# Fano resonance and spectrally modified photoluminescence enhancement in monolayer MoS$_2$ integrated with plasmonic nanoantenna array

Bumsu Lee[1†], Joohee Park[1†], Gang Hee Han[2], Ho-Seok Ee[1], Carl H. Naylor[2], Wenjing Liu[1], A.T. Charlie Johnson[2] and Ritesh Agarwal[1*]

[1]Department of Materials Science and Engineering, University of Pennsylvania, Philadelphia, PA 19104, USA

[2]Department of Physics and Astronomy, University of Pennsylvania, Philadelphia, PA 19104, USA

[†] These authors contributed equally to this work.
[*] To whom correspondence should be addressed. E-mail: riteshag@seas.upenn.edu

The manipulation of light-matter interactions in two-dimensional atomically thin crystals is critical for obtaining new optoelectronic functionalities in these strongly confined materials. Here, by integrating chemically grown monolayers of MoS$_2$ with a silver-bowtie nanoantenna array supporting narrow surface-lattice plasmonic resonances, a unique two-dimensional optical system has been achieved. The enhanced exciton-plasmon coupling enables profound changes in the emission and excitation processes leading to spectrally tunable, large photoluminescence enhancement as well as surface-enhanced Raman scattering at room temperature. Furthermore, at low temperatures, due to the **decreased damping of $MoS_2$ excitons interacting with the plasmonic resonances of the bowtie array, stronger exciton-plasmon coupling is achieved resulting in a Fano lineshape in the reflection spectrum. The Fano lineshape, which is due to the interference between the pathways involving the excitation of the exciton and plasmon, can be tuned by altering the coupling strengths between the two systems via changing the design of the bowties lattice. The ability to manipulate the optical properties of two-dimensional systems with tunable plasmonic resonators offers a new platform for the design of novel optical devices with precisely tailored responses.**

Two-dimensional crystals such as transition metal dichalcogenides (TMDs) offer a unique material platform to investigate novel properties of strongly confined systems that can be further manipulated with heterogeneous integration with other low-dimensional systems[1-5]. Due to weak interlayer bonds, TMDs can be easily layered in two-dimensions to assemble atomically thin crystals, properties of which are being investigated for potential applications such as ultra-thin light-emitting diodes, photodetectors, transistors, and biosensors[6]. Molybdenum disulphide ($MoS_2$) is one such semiconductor, which transitions from an indirect to a direct electronic band gap material when it becomes single-layered[7,8]. Broken inversion symmetry and strong spin-orbit coupling in a monolayer of $MoS_2$ result in interesting properties such as valley polarization, optical nonlinearity and strong exciton binding energy[9-11]. However, their applications in fabricating photonic devices can be limited due to large nonradiative decay rates and small energy difference between different valleys with direct and indirect energy gaps, leading to extremely low photoluminescence (PL) quantum efficiency. Although increase in PL quantum efficiency of $MoS_2$ has been achieved via chemical functionalization[12] and integration with photonic crystal cavities[13], the reported enhancements have been low and also requiring large device footprints. Therefore, in addition to achieving large enhancements in PL, it would also be desirable if the light-matter coupling in these systems can be tuned from weak to strong coupling limit, to enable a whole new suite of optoelectronic applications with precisely tailored responses.

Metal nanostructures upon interacting with light produce collective charge oscillations, known as localized surface plasmons (LSPs), which induce strong resonant fields in their vicinity[14]. In bowtie nano-resonators, the paired metal triangles build up charges at the apexes, analogous to a capacitor due to the lightening rod effect, leading to very strong LSP modes in comparison to other geometries[15,16]. When plasmonic nanostructures interact with an emitter, the optical properties of the emitter can be manipulated[16], and depending on the optical field in the system and its interaction with the active medium, weak- or strong-coupling regimes can be reached[17]. In the weak-coupling regime, the fluorescence of the emitter is modified by its optical environment, which leads to changes in the frequency-dependent spontaneous emission rate, known as the Purcell effect[17,18]. In the strong-coupling regime, the emitter and the optical cavity form a hybrid system by coherently exchanging energy, resulting in a drastically different energy spectrum and properties[17].

Here, we report a unique 2-D exciton-plasmon system composed of a monolayer of $MoS_2$ integrated with a planar silver bowtie array to significantly tailor light-matter interactions in this system. Periodic patterning of bowties into an array leads to LSP resonance linewidths much narrower than that of a single bowtie because of Rayleigh-Wood anomaly arising from the diffraction condition of the pattern[17,19-21]. By changing the geometrical parameters of the bowtie array, we achieve broad tunability of the spectral positions and linewidths of lattice-coupled LSP (lattice-LSP) modes. Through coupling of $MoS_2$ with the lattice-LSP mode, PL and Raman scattering of $MoS_2$ are drastically modified at room temperature. At low temperatures, due to the lower dephasing rate of excitons, much stronger exciton-plasmon coupling is realized, which is manifested as Fano resonances in the reflection spectra. These results demonstrate tunable optical modulation of the 2-D active medium induced by the plasmonic nanoresonators, leading to improvements in emission intensity and new optical properties based on interference between different excitation pathways of the exciton-plasmon system.

$MoS_2$ flakes were grown on $SiO_2$/Si substrates by chemical vapor deposition (CVD)[22,23]. 50 nm-thick silver bowtie arrays with varying geometrical factors were patterned directly on the $MoS_2$ flakes via electron-beam lithography (see Methods and Fig. 1a,b). Silver was used due to its strong plasmonic resonances as well as relatively low dissipation in the visible frequency range. Raman and PL measurements were performed at room temperature using the 532 nm line of an Argon ion laser focused to a spot size of 1 $\mu$m (see Methods). The Raman spectrum of the bare $MoS_2$ sample displays modes at 384 cm$^{-1}$ and 403 cm$^{-1}$ corresponding to the in-plane ($E^1_{2g}$) and out-of-plane ($A_{1g}$) modes respectively (Fig. 1c, black curve), that are fingerprints of a monolayer of $MoS_2$ (ref. 24). Similarly, PL spectrum from bare $MoS_2$ (Fig. 1d, black curve) displays peaks corresponding to A- and B-excitons at ~1.9 eV and 2.0 eV, respectively, that arise from the direct band gap and strong spin-orbit coupling in monolayer $MoS_2$ (ref. 7,8). Although the PL spectrum from the monolayer of $MoS_2$ shows excitonic features at room temperature, the emission is not strong because of the low intrinsic emission efficiency in these systems, consistent with previous studies[13,25].

However, both the Raman scattering and PL from the monolayer $MoS_2$ integrated with silver bowtie array (refer to the figure captions for the bowtie geometrical factors) display enhancements of more than an order of magnitude compared to bare $MoS_2$ (Fig. 1c,d, red curves).

Typically, the mechanisms behind enhanced Raman scattering and PL emission are surface-enhanced Raman scattering (SERS) and surface-enhanced fluorescence (SEF) respectively, as observed in molecular systems placed in the vicinity of plasmonic nanostructures[26,27]. The enhanced emission at room temperature is attributed to the Purcell effect due to the local field increase at the position of the emitter arising from the lattice-LSP resonances of the bowtie array, suggesting weak coupling between the $MoS_2$ excitons and lattice-LSPs[18].

In order to study the lattice-LSP resonances of the bowtie array, numerical calculations were carried out by finite-difference time-domain (FDTD) method (see Methods and Supplementary Information). The calculated extinction cross-sections of a single bowtie (gap separation (g) = 20 nm, size of triangle (s) = 100 nm) on a 300 nm $SiO_2$/Si substrate for both TE (electric-field (*E*-field) parallel to the bowtie axis) and TM (*E*-field perpendicular to the bowtie axis) polarizations are shown in Fig. 1e. The TE polarized LSP mode was obtained at 2.16 eV while the TM polarized LSP mode was located at a higher energy (2.5 eV) due to a stronger restoring force of the charge oscillation. Both LSP modes are very broad with ~0.4 eV linewidth, spanning most of the visible region of the electromagnetic spectrum. However, when the bowties are periodically patterned into a 2-D array, the plasmonic near-fields of individual bowties interact with neighboring bowties leading to collective LSP resonances from this 2-D plasmonic crystal (Supplementary Fig. S1). Coherent coupling between the bowtie's LSPs and the lattice diffraction modes produces a new type of lattice-LSP resonances[17,19-21] with narrow resonance linewidths; for example, the calculated quality factor (*Q*) of a lattice-LSP resonance at 1.9 eV (Fig. 1f, green curve) is 21, in comparison to a *Q* of 6 for a single-bowtie LSP resonance (TE, Fig. 1e). Furthermore, mode tunability of lattice-LSP resonances improves significantly compared to single bowtie LSPs, as represented in the calculated *E*-field enhancement profiles in

Fig. 1f. Spatial *E*-field profiles (insets in Fig. 1e,f) show that the plasmonic fields are mostly concentrated between the gaps or tips of the bowties with small optical mode volumes, which can explain the observed SERS and SEF.

To understand the correlation between the lattice-LSP modes and MoS$_2$ PL and Raman scattering intensity, detailed spectroscopic characterization including polarization-dependent reflectance, PL and Raman spectra were performed on individual constituents and coupled MoS$_2$-bowtie array system. To resolve the lattice-LSP modes, reflection spectra from the bowtie array patterned on 300 nm SiO$_2$/Si substrate were measured in TE and TM polarizations. Figure 2a shows good agreement between the experimental and calculated differential reflectance ($\Delta R/R = (R_{sample} - R_{background})/R_{background}$) from the bowtie array on SiO$_2$/Si substrates, which features broad dips for both polarizations. The dips in the reflection spectra are closely related to the lattice-LSP mode positions, as confirmed by the pitch-variation studies (for a detailed discussion, see Supplementary Information and Fig. S2)**;** comparison of the calculated *E*-field enhancement profiles and the corresponding $\Delta R/R$ reveals that the lattice-LSP modes are near the dips, due to the interference between a broad background reflection from a 300 nm thick SiO$_2$ layer, backscattering of LSP modes and in-plane scattering of lattice modes from the incident illumination (Fig. 2a,b). The emission from bowtie-MoS$_2$ exhibits distinct polarization dependence (Fig. 2c) and is closely related to the corresponding *E*-field enhancement (Fig. 2b). For example, in TE polarization (red curves, Fig. 2b,c), the maximum *E*-field enhancement at the A-exciton energy corresponds to enhanced A-exciton emission, while in TM polarization (blue curves, Fig. 2b,c), strong Raman scattering intensity is consistent with the maximum *E*-field enhancement near these Raman modes. These are also confirmed in the polarization-dependent 2-D intensity scans measured at the Raman and A-exciton energies (Fig. 2d,e), consistent with the simulated enhancements (Fig. 2b).

The effect of spectral tuning of lattice-LSP modes on modification of $MoS_2$ emission was systematically studied by varying the geometrical factors of the bowtie array (four representative patterns are shown in Fig. 3a, labeled (i)-(iv)). In order to obtain the lattice-LSP mode positions and correlate them to the emission profiles, $\Delta R/R$ spectra were measured for the four patterns on $SiO_2$/Si substrates (Fig. 3b). The frequency-dependent emission enhancements of the monolayer $MoS_2$ integrated with these four bowtie patterns (Fig. 3c,d) show that the increase of the emission in the bowtie-$MoS_2$ varies up to ~40X depending on the spectral positions of lattice-LSP modes. When the lattice-LSP mode of the bowtie array is in resonance with A- or B-excitons, as in pattern (iii) or (ii) respectively, the PL of the bowtie-$MoS_2$ system increases at the corresponding spectral positions (Fig. 3c,d). On the other hand, detuning of lattice-LSP mode and the A- or B-exciton resonance leads to only minor PL enhancement, as observed for pattern (i). For more data on the correlation of different lattice-LSP modes to the emission profiles, see Supplementary Fig. S3. In the enhancement spectra (Fig. 3d) obtained for these patterns, a new maximum is observed at 1.83 eV, which occurs at the previously reported trion state[25,28]. Although the origin of the enhancement of the trion state in our bowtie- $MoS_2$ system is unclear, it is possible that the excitons in $MoS_2$ interact with the metal leading to charge transfer, promoting the formation of trions.

The normalized PL spectra from the four patterns (Fig. 3e) clearly show the changes in spectral shapes, which depend on the lattice-LSP mode positions. This feature, known as spectrally modified SEF, is typically observed in molecular systems coupled with plasmonic structures[29-32]. For spectrally modified SEF, the total fluorescence enhancement is given by, $g_{total}$

= $g_{ex} \cdot g_{em}$, where $g_{ex}$ and $g_{em}$ are the excitation and emission rate enhancements, respectively. The increase in the excitation rate is due to the increased coupling between the incident light and the emitter, arising from the enhanced local field. In this case, the excitation rate enhancement depends only on the frequency of incident light ($\omega_i$), and should not affect the spectral shape. For example, the calculated $g_{ex}$ for sample (iii) (which shows maximum enhancement of the A-exciton emission) is ~1.8 (Supplementary Fig. S4), while the observed PL enhancement is much higher. This in addition to the observed spectral modification in the bowtie-$MoS_2$ system cannot be explained only by the enhancement of the excitation rate, which is not very high because the laser excitation energy (2.33 eV) is considerably far away from the lattice-LSP modes. In addition, sample (i) shows a high-energy tail above the B-exciton emission in the normalized PL spectrum (Fig. 3e, blue curve), corresponding to the lattice-LSP mode located at higher energy (Fig. 3b, blue curve), attributed to the emission from non-thermalized excitons, similar to the hot-exciton emission observed in plasmonically-coupled CdS nanowires[33].

The emission enhancement ($g_{em}$) can be explained by the change in the quantum yield of the emitter in the presence of plasmonic resonators. The emission quantum yield ($k_r/(k_r+k_{nr})$), where $k_r$ and $k_{nr}$ are the radiative and the non-radiative decay rates, respectively, changes as both $k_r$ and $k_{nr}$ change when the emitter is coupled with plasmonic resonators. The radiative decay rate is affected mainly by the Purcell effect whereas the non-radiative decay rate depends on plasmonic losses and exciton quenching. Generally, in an electronic multi-level system, $k_r$ and $k_{nr}$ are frequency dependent and depend on the competition between the intra- and inter-band relaxation rates. Therefore, our frequency-dependent PL enhancement ($g_{total}(\omega)$) can only be explained by the quantum yield increase ($g_{em}(\omega)$), which is frequency-dependent. For our bowtie-$MoS_2$ (sample (iii)), the overall emission enhancement (integrated PL) is ~25, and therefore, accounting for the excitation rate increase of 1.8, the Purcell-type emission enhancement is ~14. This significant Purcell-type enhancement indicates effective coupling between 2-D excitons and lattice-LSP resonances and is in agreement with recent theoretical studies supporting the large increase in the radiative decay rate of an emitter in the bowtie gap[34,35]. Further improvements in the emission yield can be achieved by inserting a thin dielectric layer between $MoS_2$ and metal structures to optimize the competition between exciton-quenching and plasmonic field enhancement at this interface.

In order to study how the bowtie-$MoS_2$ coupled system responds at low temperatures when the exciton-dephasing rate is reduced, the far-field $\Delta R/R$ spectra for bare $MoS_2$, bowtie array, and bowtie-$MoS_2$ system were measured at 77 K (Fig. 4a). Clear absorption dips associated with the A- and B-excitons in bare $MoS_2$ appear at 1.92 eV and 2.1 eV (black curve). $\Delta R/R$ of the bowtie array shows two broad dips (blue curve), with the lattice-LSP modes spectrally overlapping with the $MoS_2$ excitons. In the combined bowtie-$MoS_2$ system, however, an interesting phenomenon is observed; an asymmetric feature resembling Fano lineshape appears in the reflection spectra at the $MoS_2$ exciton energies (red curve). The Fano lineshape is a result of the spectral interference between a narrow discrete resonance and a broad continuum of states[36,37]. In our case, the $MoS_2$ excitons with a sharp resonance at 77 K couple with the broad continuum of plasmons. Since the linewidth of the A-exciton is much narrower than that of the B-exciton, more distinct Fano lineshape was observed at the A-exciton spectral region.

FDTD calculation simulating the $\Delta R/R$ response of the experimental system (Fig. 4b) reproduces the reflection spectra of bare $MoS_2$ (black curve) and the bowtie array (blue curve) as well as the Fano lineshape of the integrated bowtie-$MoS_2$ system (red curve). To further investigate the physical mechanism of the observed Fano features, the absorption spectra (Fig.

4c) were calculated for bare $MoS_2$, bowtie array, $MoS_2$ in the combined bowtie-$MoS_2$ system, and the bowtie-array in the combined bowtie-$MoS_2$ system. Bare $MoS_2$ shows a sharp excitonic absorption (black curve) while the bowtie array shows broad plasmon absorption (blue curve), as expected. For the coupled bowtie-$MoS_2$ system, absorption in $MoS_2$ (green curve) shows absorption enhancement in comparison to the absorption of bare $MoS_2$ due to the strong plasmonic field. However, absorption in the bowties (red curve) in the coupled system features complex interference at the $MoS_2$ exciton energies compared to a typically broad lattice-LSP absorption. This demonstrates that the absorption in the bowtie array of the combined system exhibits Fano interference due to the strong interaction between $MoS_2$ and the bowtie array.

The physical mechanism (Fig. 4b, inset) behind the measured Fano resonance in the bowtie-$MoS_2$ system is the interference that occurs predominantly in the excitation process[38-40]; in the exciton-plasmon coupled system, the exciton life time is much longer than that of plasmons and therefore the excitation rate of excitons increases without any significant changes in the spectral shape or position due to the enhanced local plasmonic field (Fig. 4c, green curve). On the other hand, absorption in the bowtie array can be significantly affected by the interference from the excitons with long lifetime via the exciton-plasmon dipole-dipole coupling. Thus, in the bowtie-$MoS_2$ system, two major optical paths exist for the excitation of plasmonic modes[39-41], i.e., direct excitation of the plasmons and their indirect excitation via the dipole-dipole coupling of $MoS_2$ excitons and plasmons. The dipole-dipole interaction increases due to the local field enhancement arising from the surface plasmons and leads to exciton-plasmon coupling beyond the perturbative regime into an intermediate state between weak (Purcell) and strong (Rabi) coupling regime. The constructive and destructive interference between the two optical pathways leads to Fano resonance measured in our reflection spectra, consistent with the large PL enhancement addressed earlier (Fig. 3).

Fano shapes vary in our experimental $\Delta R/R$ measurements depending on the linewidth of lattice-LSP mode and detuning between $MoS_2$ excitons and tunable lattice-LSP modes (Fig. 5). Three representative samples are compared to examine their Fano spectral features by controlling their geometric factors and hence their plasmonic resonances. For a very small detuning and sharper linewidth of bowtie resonances (Fig. 5a), much sharper Fano feature with more distinct asymmetry was observed in comparison to the samples with increasing detuning[38,42,43] (Fig. 5b,c). Moreover, no Fano response was observed in Fig. 5c due to almost no overlap between the exciton and lattice-LSP resonances. These observations strongly support the claim that our Fano features originate from the strong coupling between $MoS_2$ excitons and lattice-LSP modes.

Fano resonances appear in many systems ranging from autoionization of atoms to plasmonic nanostructures and metamaterials[36,37,44-48]. However, most observations are in passive systems with no involvement of electronic resonances[45-48]. Recently, Fano resonance and Rabi oscillations have been demonstrated in *J*-aggregate dye molecules coupled with metal nanostructures due to the enhanced interaction between the excitons and LSPs[41,43,49,50] The bowtie-$MoS_2$ system with very small mode volume, displays Fano resonances due to the significantly enhanced optical coupling between the 2-D $MoS_2$ excitons and bowtie LSP resonances, which can be modulated via the spectral tuning of lattice-LSP resonances.

In conclusion, significant modification of the emission properties of monolayer $MoS_2$ was demonstrated upon its integration with silver bowtie nanoantenna arrays. The strong plasmonic local field from lattice-induced LSP resonances in the bowtie arrays gives rise to enhanced Raman scattering and PL in $MoS_2$ at room temperature. Depending on detuning of $MoS_2$ exciton and lattice-LSP mode, spectrally modified PL enhancement was exhibited, arising from the Purcell effect. At low temperature, due to the strong dipole-dipole interaction between $MoS_2$ excitons and lattice-LSP modes with sharper resonances, quantum interference arises in the excitation process and is manifested as Fano-like asymmetric reflection spectra. Since the Fano resonance lineshape and spectral position is very sensitive to local perturbations, it can be utilized to assemble optical switches and sensors[44]. Tailoring light-matter interactions between atomically thin semiconductor crystals and plasmonic nanostructures as demonstrated in this work will be critical to realize new physical phenomena and fabricate novel optical devices with applications ranging from improved light sources, detectors, and sensors to photovoltaics.

## Methods

**Device fabrication.** Monolayers of $MoS_2$ were grown on $SiO_2$/Si substrate (300 nm thermally grown $SiO_2$) via chemical vapor deposition as reported earlier[22,23]. Bowtie structures were patterned on the as-grown $MoS_2$ substrate by electron beam lithography, followed by the deposition of 50 nm silver by electron-beam evaporation.

**Optical measurements.** Photoluminescence and Raman measurements were performed on the NTEGRA Spectra Probe system equipped with a 0.7 NA objective (~400 nm spatial resolution). A c.w. 532 nm wavelength laser focused to a spot size of 1 $\mu$m with an excitation power of 0.16 mW was used for all optical measurements. For reflectance measurements, the sample was excited by a white light source in our home-built optical microscopy setup with a 60X (0.7 NA) objective. The samples were loaded in an optical microscopy cryostat (Janis ST-500) and cooled to 77 K with liquid nitrogen.

**Numerical calculations.** Reflectance, $E$-field enhancement, and absorption spectra for the silver bowtie plasmonic nanocavities on $SiO_2$/Si were calculated via three-dimensional finite-difference time-domain (FDTD) simulations using a vertically incident broadband plane-wave source. For detailed information, see Supplementary Information.

**Acknowledgements**

This work was supported by the US Army Research Office (grant numbers W911NF-09-1-0477 and W911NF-11-1-0024) and the National Institutes of Health through the NIH Director's New Innovator Award Program (1-DP2-7251-01). GHH, CHN, and ATCJ recognize support form the National Science Foundation Accelerating Innovation in Research program AIR ENG-1312202 and the Nano/Bio Interface Center NSF NSEC DMR08-32802. R.A. and A.T.C.J acknowledge Seed Project support from the LRSM, NSF MRSEC DMR-1120901.

**Competing financial interests:** The authors declare no completing financial interests.

**Figure legends**

**Figure 1 | Optical properties of monolayer MoS$_2$-bowtie resonator array system at room-temperature**. **a**, Scanning electron microscope (SEM) image showing the silver bowtie array directly patterned on well-defined, stacked triangular flakes of mono- and bi-layer MoS$_2$. Larger triangular flake of darker contrast corresponds to a single layer and smaller flake of lighter contrast to a bilayer. **b**, Device schematic indicating the geometrical factors of a bowtie array: gap separation (g), thickness of the metal deposition (h), side length of a triangle (s), and unit cell dimension or pitch (p = (p$_x$, p$_y$)). **c**, Raman scattering spectra of bare MoS$_2$, bowtie array, and bowtie-MoS$_2$. Geometrical factors: g = 20 nm; h = 50 nm; s = 100 nm; p = (500 nm, 300 nm). For all bowtie arrays studied throughout the paper, g = 20 nm and h = 50 nm. **d**, PL spectra of bare MoS$_2$, bowtie array, and bowtie-MoS$_2$. Inset shows PL in log scale. **e**, Calculated extinction of a single bowtie (g = 20 nm; h = 50 nm; s = 100 nm) with the inset displaying the *E*-field intensity profiles for TE and TM polarizations with respect to the long axis of a bowtie. **f**, Calculated *E*-field enhancements in the array of bowties (s = 100 nm) with four different unit cell lengths (p). Inset shows the *E*-field intensity profiles for TE or TM polarizations.

**Figure 2 | Polarization-dependent reflection spectra of the bowtie resonator array and photoluminescence spectra of MoS$_2$ coupled with bowtie resonator array.** **a**, Normalized differential reflectivity (Δ*R/R*) spectra of experimental and calculated lattice-LSP modes in TE and TM polarizations for bowtie array on SiO$_2$/Si substrate. Geometrical factors: s = 100 nm, p = (500 nm, 300 nm). **b**, Average *E*-field enhancements for bowtie array on SiO$_2$/Si calculated for the total area of the unit cell **c**, Experimental emission (PL and Raman) spectra for bowtie-MoS$_2$

system. Inset; zoomed-in spectra near the Raman active region. **d**, Optical microscope image of bowtie arrays patterned in two orthogonal orientations on a MoS$_2$ flake. **e**, 2-D intensity scans measured at the Raman mode and A-exciton energies. Scanning area indicated in **d**. Data in **e** shows different polarization response for Raman (horizontally patterned bowties) and A-exciton emission (vertically patterned bowties).

**Figure 3 | Spectral modification of MoS$_2$ photoluminescence coupled with bowtie resonator arrays with different lattice-LSP dipole resonances at room temperature. a**, SEM images of four bowtie-array samples with different bowtie sizes and pitch values: (i) s = 100 nm, p = (400 nm, 500 nm) (ii) s = 100 nm, p = (400 nm, 300 nm) (iii) s = 100 nm, p = (300 nm, 200 nm) (iv) s = 170 nm, p = (500 nm, 800 nm). **b**, $\Delta R/R$ spectra associated with the lattice-LSP modes of the four different bowtie patterns on SiO$_2$/Si substrates. **c**, PL spectra of bare MoS$_2$ (black) and four different patterns. **d**, Wavelength-dependent PL enhancements (ratios of PL obtained for bowtie-MoS$_2$ sample to PL of bare MoS$_2$) for the four patterns on monolayer MoS$_2$. Inset shows the enhancement of A-exciton emission for all four patterns. **e**, Normalized PL spectra of bare MoS$_2$ (black) and the four different bowtie patterns on MoS$_2$.

**Figure 4 | Fano resonance in the reflection spectrum of bowtie-MoS$_2$ at 77 K.** Bowtie geometry: s = 100 nm, p = (300 nm, 200 nm). **a** and **b**. Experimental and calculated $\Delta R/R$ spectra for bare MoS$_2$, bowtie array, and bowtie-MoS$_2$ system. Clear Fano resonances are observed at the A- and B-exciton spectral region. Inset in **b** shows the schematic for the mechanism of observed Fano resonance in the exciton-plasmon system. **c**, Calculated absorption spectra for the

bare MoS$_2$ (black), bowtie array (blue), MoS$_2$ in the combined bowtie-MoS$_2$ system (green), and the bowtie in the combined bowtie-MoS$_2$ system (red). The dotted circle highlights the region of MoS$_2$ excitons where Fano resonances are observed in the refection spectra in **a**.

**Figure 5 | Controlling Fano spectra of the bowtie-MoS$_2$ system by tuning the lattice-LSP modes to overlap with MoS$_2$ excitons at 77 K. a-c**, Experimental $\Delta R/R$ spectra of bare MoS$_2$, bowtie, and bowtie-MoS$_2$ system for three different samples with the following excitation polarization and geometrical factors. TE polarization, s = 100 nm, p = (400 nm, 300 nm) (**a**), TM, s = 140 nm, p = (700 nm, 400 nm) (**b**), TE, s = 100 nm, p = (800 nm, 700 nm) (**c**). Clear Fano resonances are observed when the bowtie lattice-LSP modes overlap with MoS$_2$ excitons.

# Figures

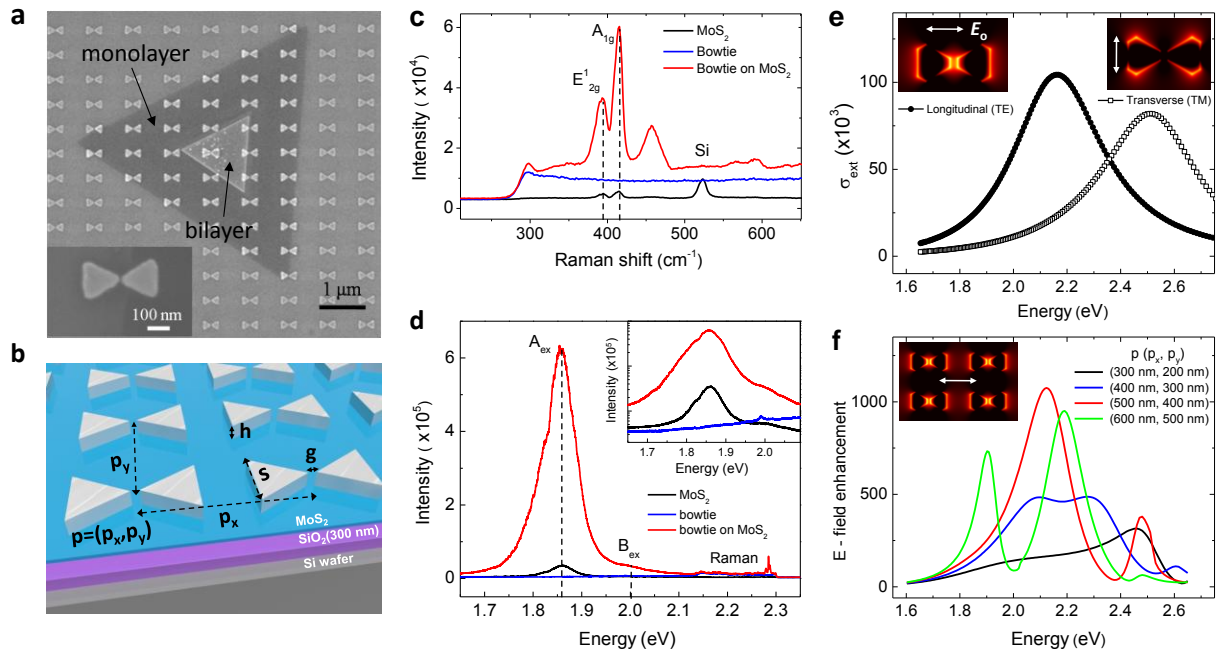

**Figure 1**

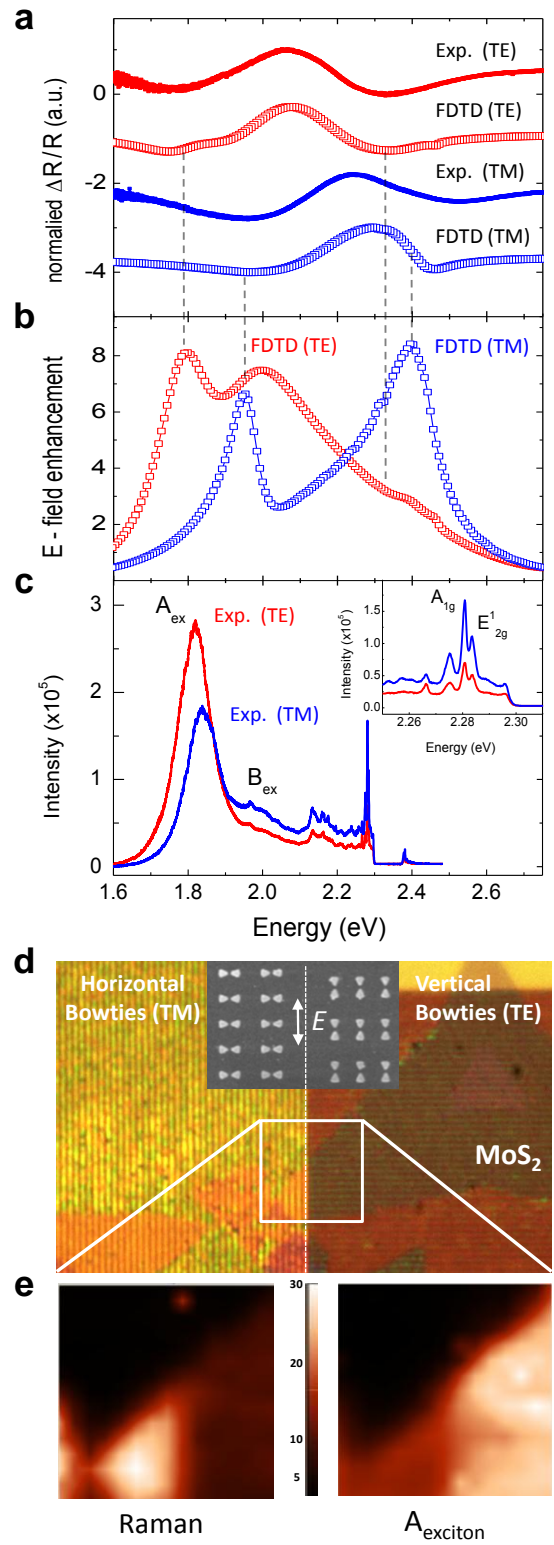

**Figure 2**

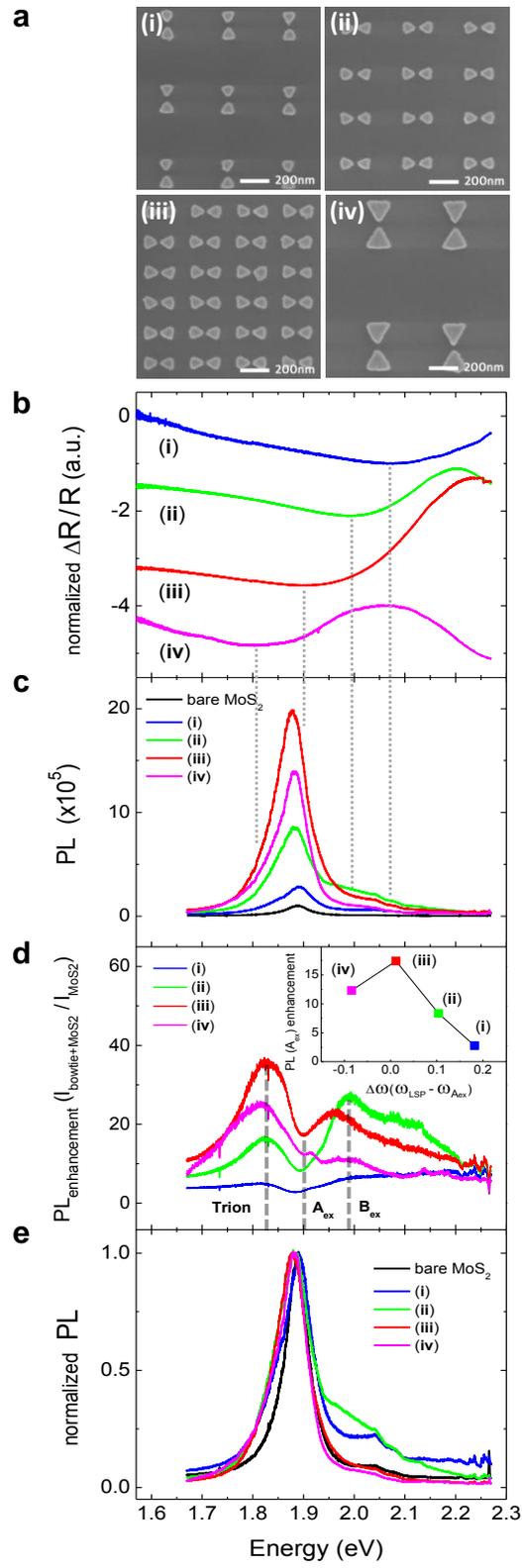

**Figure 3**

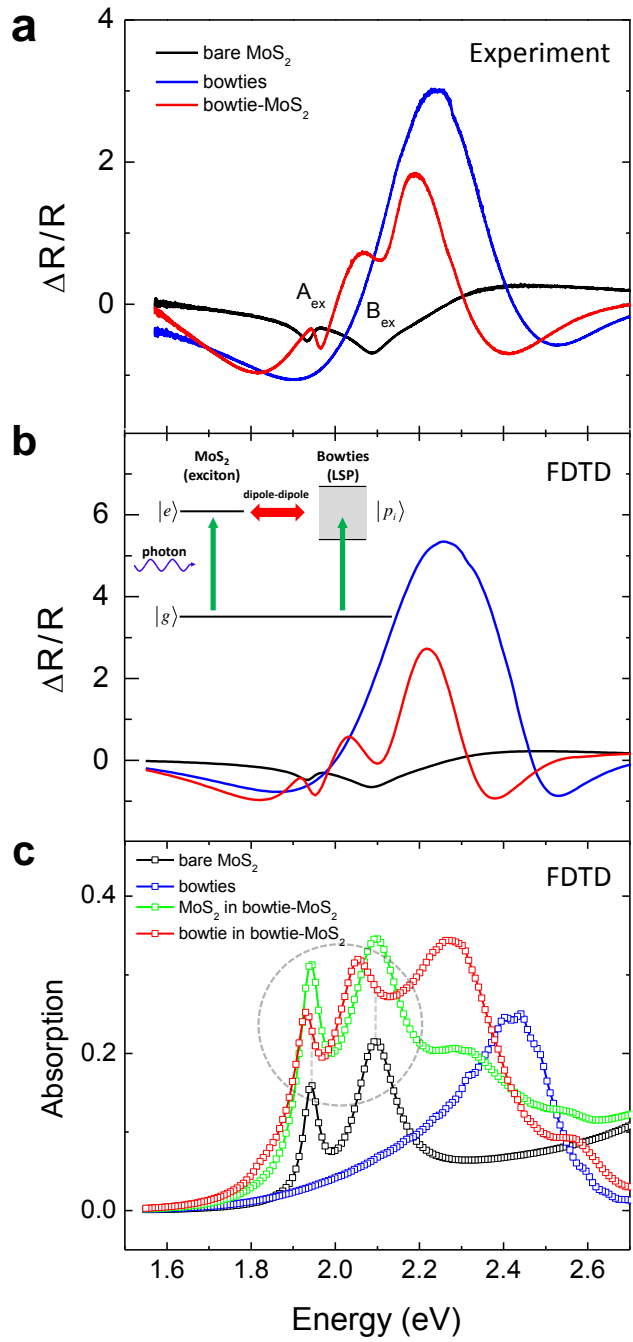

**Figure 4**

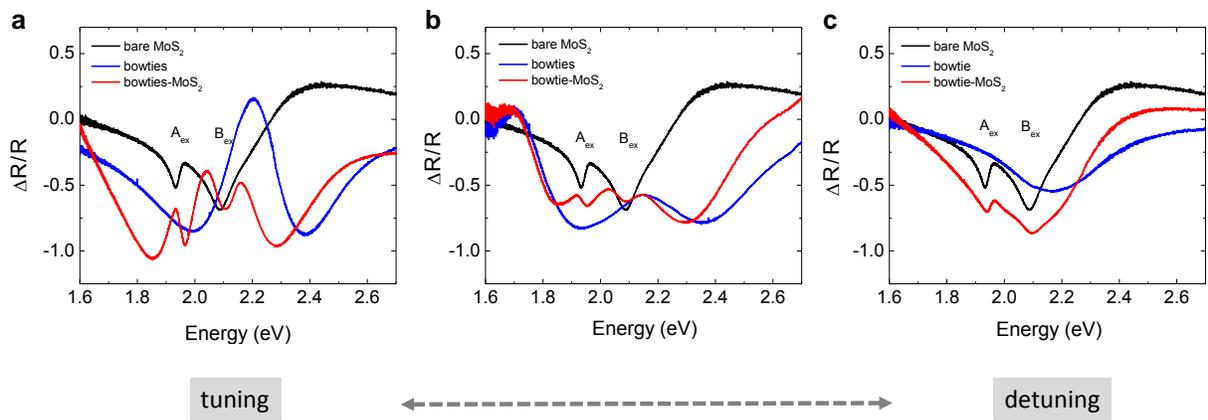

**Figure 5**

# Supplementary Information

# Fano resonance and spectrally modified photoluminescence enhancement in monolayer $MoS_2$ integrated with plasmonic nanoantenna array

Bumsu Lee, Joohee Park, Gang Hee Han, Ho-Seok Ee, Carl H. Naylor, Wenjing Liu, A.T. Charlie Johnson and Ritesh Agarwal

## 1. LSP of a single bowtie vs. Lattice-LSP of bowtie arrays

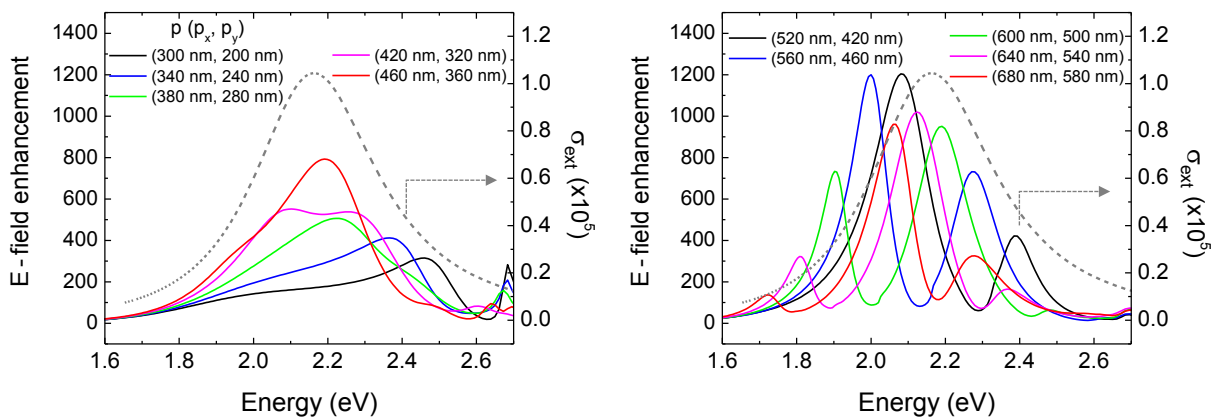

**Figure S1** | Calculated *E*-field enhancements of Ag bowtie arrays with varying pitch (p) values (solid curves), compared with the calculated extinction cross-section of a single bowtie (dotted curve) for TE polarization excitation. Bowtie geometry: gap separation (g) = 20 nm, thickness of the metal deposition (h) = 50 nm, side length of a triangle (s) = 100 nm.

The coherent coupling between the single bowtie's LSPs and the bowtie array's lattice diffraction modes results in lattice-LSP resonances. The calculated extinction cross-section of a single bowtie (g = 20 nm, s = 100 nm) on a 300 nm $SiO_2$/Si substrate for longitudinal (TE) polarization shows a broad LSP mode centered at 2.16 eV (dotted gray curve), which is compared to the calculated *E*-field enhancement profiles of 10 bowtie arrays with varying pitch showing the evolution of the lattice-LSP modes (solid curves) for TE polarization excitation. The *E*-field enhancements are calculated from the center of the bowtie's gaps. It is shown that the linewidth of lattice-LSP modes are sharper than that of single bowtie LSP resonance.

## 2. *E*-field enhancement of lattice-LSP modes vs. Reflectance

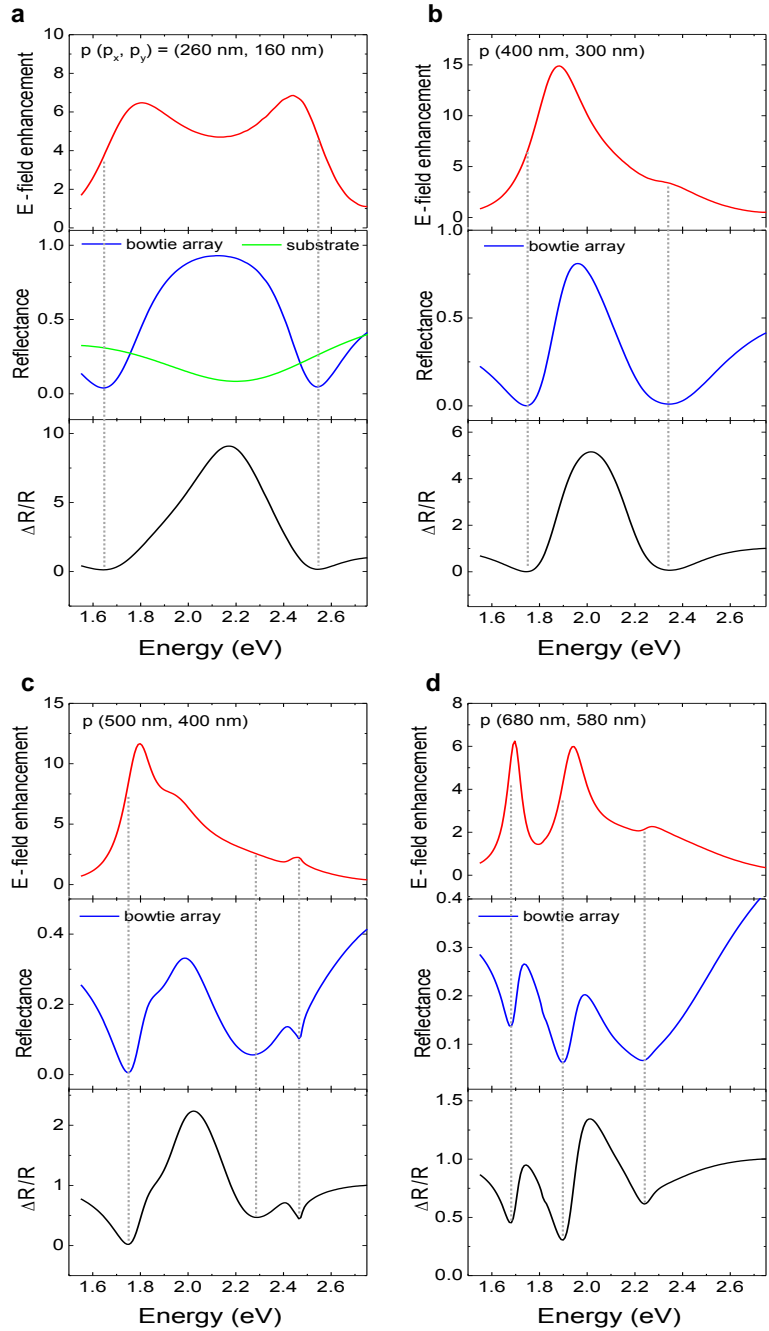

**Figure S2** | Calculated average *E*-field enhancement, reflectance (ratio of reflected light to incident light intensity), and Δ*R/R* ((*R*$_{sample}$ − *R*$_{background}$)/*R*$_{background}$)) spectra for the bowtie arrays

on SiO$_2$/Si substrate with different pitch values. Δ*R/R* shows enhanced features due to subtraction of the background. Bowtie geometrical factors: g = 10 nm, h = 50 nm, s = 100 nm.

The *E*-field enhancement in bowtie arrays (Fig. S2) explains the location of lattice-LSP modes (red curves). Since lattice-LSP modes are a coupled system between individual bowtie LSPs and lattice modes, the reflection spectra are affected by both the superradiant LSP dipole modes and subradiant lattice (grating) modes due to radiation progress in the plane of the bowtie array for normal excitation. In addition, a 300 nm thick SiO$_2$/Si substrate produces a broad reflection signal as a background (Fig. S2a, green curve). Therefore, the reflection spectra of the bowtie arrays are mainly composed of all these factors: the broad reflectance of 300 nm thick SiO$_2$/Si substrate, the backscattered signal from superradiant LSP dipole mode and surface-propagating signals from the lattice modes. These signals interfere and result in the complex reflectance and Δ*R/R* spectra as shown in Fig. S2 (blue and black curves). The peaks of *E*-field enhancement spectra are located near the dips of the reflectance spectra. The lattice-LSP mode positions are thus near the dips in the reflection spectra, which is consistent in all four samples with different pitch values. In the bowtie-MoS$_2$ system, we expect that lattice-LSP mode to be slightly red-shifted because of a relatively high refractive index (*n* ~6) of the MoS$_2$ layer.

## 3. Spectrally modified SEF enhancement vs. Reflectance

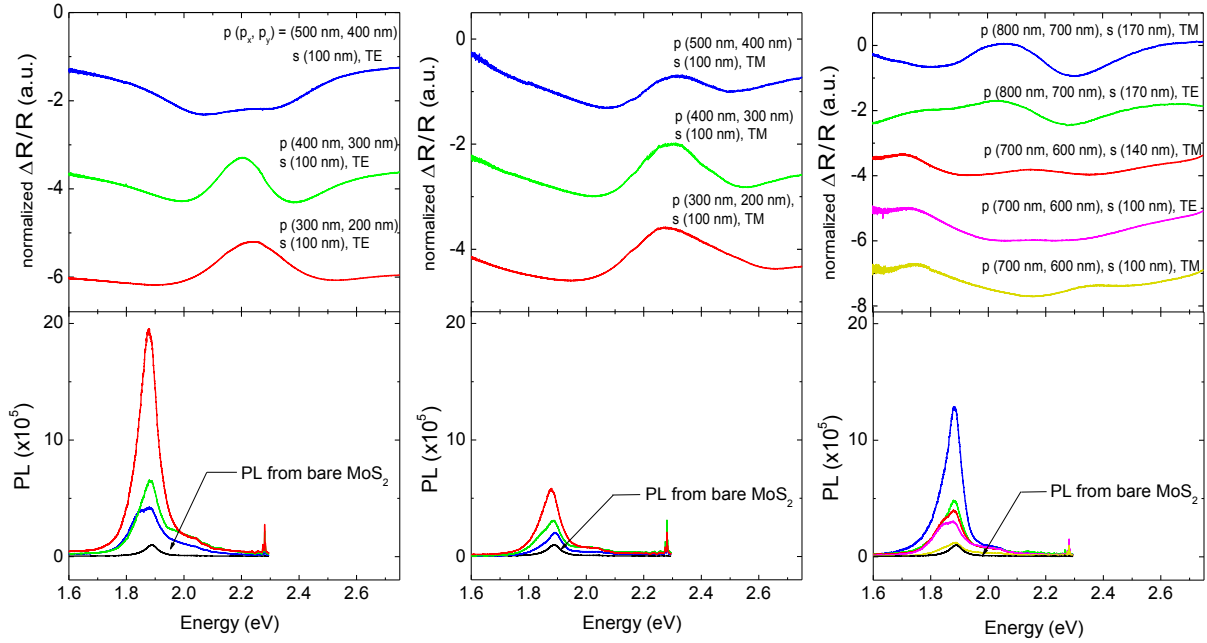

**Figure S3** | Lattice-LSP mode dependent spectrally modified PL enhancements of MoS$_2$ coupled to Ag bowtie arrays. Top panel shows the Δ*R/R* spectra of bowtie arrays with different geometrical factors and polarization, as noted above each spectrum. Bottom panel shows PL spectra of bowtie-MoS$_2$ samples from the corresponding bowtie array.

Since the lattice-LSP modes are located at around the dips of the Δ*R/R* spectra, the PL enhancement in the bowtie-MoS$_2$ system is correlated to Δ*R/R* spectra of the bowtie arrays. By varying the array pitch and/or the side length of bowties as well as the polarization of incident light, the lattice-LSP resonance can be modified, as shown in different Δ*R/R* spectra in Fig. S3. Correlating the Δ*R/R* spectra of the bowtie array to the corresponding PL spectra of the bowtie-MoS$_2$ samples shows, as expected, that PL intensity increases the most when the dip position in $\Delta R/R$ is closest to the A- or B-exciton energy.

## 4. Excitation rate ($g_{ex}$) increase in MoS$_2$ upon integration with the bowtie array at room temperature

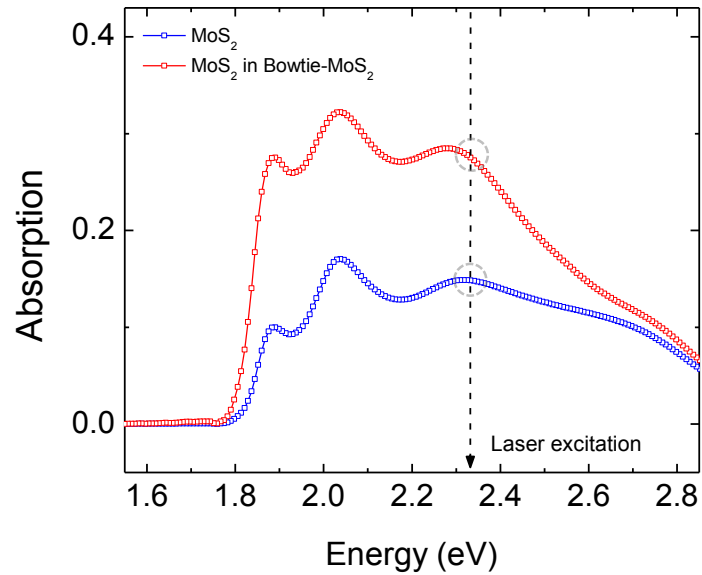

**Figure S4 |** Calculated absorption spectra for bare MoS$_2$ monolayer and bowtie-MoS$_2$ system at room temperature. The calculated results are shown for sample (iii) in Fig. 3c of the main paper. Bowtie geometrical factors: g = 20 nm, h = 50 nm, s = 100 nm, and p = (300 nm, 200 nm).

Absorption spectra were numerically calculated for bare MoS$_2$ before and after its integration with a bowtie array (Fig. S4). For bare MoS$_2$, clear A- and B-exciton peaks are observed in the calculated absorption spectrum (blue curve). For the bowtie- MoS$_2$ system, the overall absorption in MoS$_2$ increases (red curve). Absorption increases in the coupled system by a factor of 1.8 at the laser excitation energy (2.33 eV/532 nm) as indicated by the dotted circles. Therefore, the calculated excitation rate increase, $g_{ex}$, is expected to be 1.8 for this sample geometry.

## 5. FDTD calculation methods and parameters

Three-dimensional finite-difference time-domain (FDTD) simulations were performed to understand the optical properties of the Ag bowtie plasmonic nanoantenna structures on the $SiO_2$/Si substrate with and without $MoS_2$. The spatial grid sizes in $x$, $y$, $z$ directions and temporal step size were set to 2.5 nm, 1.44 nm, 0.65 nm and 0.5 $c^{-1}$ nm, respectively, where $c$ is the speed of light in vacuum. In simulations of a single bowtie, the domain size was set to 500×400×800 $nm^3$ and perfectly matched layers (PMLs) were introduced in every six boundaries to truncate the calculation domain. In simulations of a bowtie array, the domain size was set to $p_x \times p_y \times$ 800 nm. PMLs were introduced in upper and lower boundaries and periodic boundary conditions were applied in $x$ and $y$ directions. The refractive index of $SiO_2$ was set to $n = 1.46$ as in ref. 1. In simulations with $MoS_2$, the thickness of $MoS_2$ was set to 0.65 nm. A vertically incident broadband planewave source was employed by using total-field scattered-field technique[2].

The dispersive permittivities of materials were handled by auxiliary differential equation method using Drude, Lorentz, or critical-points model[3]. For Ag, Drude model, $\varepsilon(\omega) = \varepsilon_\infty - \omega_p^2/(\omega^2 + i\gamma\omega)$, was used and the parameters were $\varepsilon_\infty = 4.08849$, $\omega_p = 9.24575$ $\hbar^{-1}$ eV, and $\gamma = 0.0339505$ $\hbar^{-1}$ eV, where $\hbar^{-1}$ is the reduced Planck constant. Those were obtained by fitting measured ($n$, $k$) data of Ag in ref. 4 over the wavelength range of 400 – 800 nm. For Si, Lorentz model, $\varepsilon(\omega) = \varepsilon_\infty - \Delta\varepsilon\,\omega_0^2/(\omega_0^2 - \omega^2 - i\gamma\omega)$ was used and the parameters were $\varepsilon_\infty = 5.50461$, $\Delta\varepsilon = 6.75689$, $\omega_0 = 3.59915$ $\hbar^{-1}$ eV and $\gamma = 0.168736$ $\hbar^{-1}$ eV. Those were obtained by fitting measured ($n$, $k$) data of Si in ref. 1 over the wavelength range of 400 – 800 nm. For low temperature MoS$_2$, a model with three Lorentz oscillators, $\varepsilon(\omega) = \varepsilon_\infty - \Sigma_{p=1,2,3} f_p \omega_p^2/(\omega_p^2 - \omega^2 - i\gamma_p\omega)$ was used and the parameters were $\varepsilon_\infty = 1.03203$, $f_1 = 0.422392$, $\omega_1 = 1.94209$ $\hbar^{-1}$ eV, $\gamma_1 = 0.0404628$ $\hbar^{-1}$ eV, $f_2 = 1.01338$, $\omega_2 = 2.09462$ $\hbar^{-1}$ eV, $\gamma_2 = 0.10405$ $\hbar^{-1}$ eV, $\Delta\varepsilon_3 = 8.23947$, $\omega_3 = 2.83897$ $\hbar^{-1}$ eV and $\gamma_3 = 0.222856$ $\hbar^{-1}$ eV. Those are obtained by fitting the measured differential reflectance data of bare MoS$_2$ on SiO$_2$/Si substrate obtained at 77 K (Fig. 4A) over the wavelength range of 450 – 750 nm with an analytical solution of differential reflectance obtained by the dispersion model and transfer-matrix-method[5]. The permittivity function of Si used in FDTD simulations was also used in analytical calculation of the differential reflectance and the thickness of SiO$_2$ layer obtained from the fitting was 290 nm. For room temperature MoS$_2$, a critical-points model with five poles, $\varepsilon(\omega) = \varepsilon_\infty - \Sigma_{p=1,2,3,4,5}[(f_p + ig_p) \omega_p/(\omega_p + i\gamma_p + \omega) + (f_p - ig_p) \omega_p/(\omega_p - i\gamma_p - \omega)]$ was used and the parameters were $\varepsilon_\infty = 1$, $f_1 = 0.2968$, $g_1 = 0.20536$, $\omega_1 = 1.86599$ $\hbar^{-1}$ eV, $\gamma_1 = 0.0428117$ $\hbar^{-1}$ eV, $f_2 = 0.481256$, $g_2 = 0.220989$, $\omega_2 = 2.01418$ $\hbar^{-1}$ eV, $\gamma_2 = 0.0752466$ $\hbar^{-1}$ eV, $f_3 = 0.166503$, $g_3 = 0.244162$, $\omega_3 = 2.26616$ $\hbar^{-1}$ eV, $\gamma_3 = 0.139213$ $\hbar^{-1}$ eV, $f_4 = 3.67522$, $g_4 = 1.00497$, $\omega_4 = 2.84172$ $\hbar^{-1}$ eV, $\gamma_4 = 0.239602$ $\hbar^{-1}$ eV, $f_5 = 2.38011$, $g_5 = -1.12111$, $\omega_5 = 3.77972$ $\hbar^{-1}$ eV and $\gamma_5 = 0.752014$ $\hbar^{-1}$ eV. Those were obtained by fitting measured ($n$, $k$) data of MoS$_2$ in ref. 6.